\documentclass{JAC2003}

\usepackage{graphicx}
\usepackage{booktabs}
\usepackage{multicol}
\usepackage{amsmath}
\usepackage{url}
\begin{document}
\title{Vortex penetration field of the multilayer coating model}

\author{Takayuki Kubo\thanks{kubotaka@post.kek.jp}, Takayuki Saeki,  
High Energy Accelerator Research Organization, KEK\\ 1-1 Oho, Tsukuba, Ibaraki 305-0801 Japan\\
Yoshihisa Iwashita, Institute for Chemical Research, Kyoto University, Uji, Kyoto 611-0011, Japan}

\maketitle

\begin{abstract}
The vortex penetration field of the multilayer coating model with a single superconductor layer and a single insulator layer formed on a bulk superconductor are derived. 
The same formula can be applied to a model with a superconductor layer formed on a bulk superconductor without an insulator layer. 
\end{abstract}

%%%%%%%%%%%%%%%%%%%%%%
%%%%%%%%%%%%%%%%%%%%%%
\section{Introduction}\label{section:introduction}
%%%%%%%%%%%%%%%%%%%%%%
%%%%%%%%%%%%%%%%%%%%%%

The multilayer coating model~\cite{gurevich} was proposed as a novel method to push up the rf-breakdown field of superconducting cavities.  
The model consists of alternating layers of superconductor layers ($\mathcal{S}$) and insulator layers ($\mathcal{I}$) formed on a bulk Nb. 
The $\mathcal{S}$ layers are assumed to withstand higher field than the bulk Nb and shield the bulk Nb from the applied rf surface field $B_0$, 
by which $B_0$ is decreased down to $B_i\,(< B_0)$ on the surface of the bulk Nb.  
Then the cavity with the multilayered structure is thought to withstand a higher field than the Nb cavity,   
if $B_0$ is smaller than the vortex penetration field~\cite{beanlivingston} of the top $\mathcal{S}$ layer $B_v$, and 
$B_i$ is smaller than $\simeq 200\,{\rm mT}$, which is thought to be the maximum field for the bulk Nb.

In order to evaluate the maximum surface field of a cavity with the multilayered structure, 
correct formulae that can describe the shielded magnetic field $B_i$ and the vortex penetration field of the top $\mathcal{S}$ layer $B_v$ should be derived~\cite{kis}. 
The detailed derivation process of the magnetic field attenuation formulae and thus the shielded magnetic field $B_i$ of 
the multilayer coating model with a single superconductor layer and a single insulator layer formed on a bulk superconductor is found in Ref.~\cite{ipac13kis}, 
where the Maxwell equations and the London equations are solved with appropriate boundary conditions.

The vortex penetration field is derived from competing forces acting on a vortex at a top of the ${\mathcal{S}}$ layer. 
The vortex feel two distinct forces: 
(i) a force from an image current $j_{\rm I}$ due to an image antivortex, and
(ii) from a Meissner current $j_{\rm M}$ due to an external field.  
In this paper these two forces are evaluated, and the vortex penetration field of the multilayer coating model is derived.

%%%%%%%%%%%%%%%%%%%%%%%%%%%%%%%%%%%%%%%%%%%%%%%%%%%%%
\section{Vortex penetration field of the film superconductor in an uniform magnetic field} 
%%%%%%%%%%%%%%%%%%%%%%%%%%%%%%%%%%%%%%%%%%%%%%%%%%%%%

%
\begin{figure}[tb]
   \begin{center}
   \includegraphics[width=0.6\linewidth]{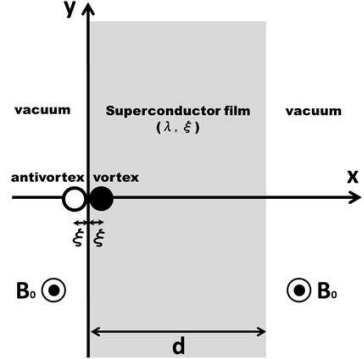}
   \end{center}\vspace{-0.6cm}
   \caption{
A superconductor film in an uniform magnetic field. 
The filled circle shows a vortex, and the open circle shows an image antivortex introduced to satisfy the boundary condition of zero current normal to the surface. 
   }\label{figure1}
\end{figure}

In this section the vortex penetration field of the superconductor film in an uniform magnetic field is evaluated as a simple example. 
The procedures shown here can be applied to the multilayer coating model in the next section.

Let us consider a superconductor film immersed in an uniform magnetic field (see Fig.~\ref{figure1}).   
The film is parallel to the $y$-$z$ plane and thus parpendicular to the $x$-axis. 
The region $0 \le x \le d$ is the superconductor with the London penetration depth $\lambda$ and the coherence length $\xi$, and
the other regions $x<0$ and $x>d$ are vacuum.  
The applied magnetic field is parallel to the $z$-axis, and is given by ${\bf B}_{\rm ext}=(0,\,0,\,B_0)$.   
It is assumed that the material of the film is an extreme Type II superconductor ($\lambda \gg \xi$), 
and the film thickness is larger than the the coherence length ($d \gg \xi$).

Suppose that a vortex that is parallel to the $z$-axis is at $(x,y)=(\xi,0)$, an edge of the film. 
Then an antivortex at $(x,y)=(-\xi,0)$ is introduced as an image of the vortex to satisfy the boundary condition of zero current normal to the surface as shown in Fig.~\ref{figure1}. 
The magnetic field of the antivortex is given by ${\bf B}_{\rm I} =(0,0, (-\phi_0/2\pi\lambda^2)\ln(\lambda/r))$ for $\xi < r < \lambda$~\cite{tinkham}, 
where $\phi_0=2.07\times10^{-15}{\rm Wb}$ is the flux quantum and $r$ is a distance from its core at $(x,y)=(-\xi,0)$. 
The associated current density at the vortex position $(x,y)=(\xi,0)$ is given by ${\bf j}_{\rm I}=(0,j_{{\rm I}\, y},0)$, 
where $j_{{\rm I}\, y} = -\phi_0/(4\pi\mu_0 \lambda^2 \xi)$. 
Then the vortex receives the Lorentz force~\cite{tinkham}: 
\begin{eqnarray}
{\bf f}_{\rm I} = {\bf j}_{\rm I}\times \phi_0 {\bf \hat{z}} = - \frac{\phi_0^2}{4\pi\mu_0 \lambda^2 \xi} {\bf \hat{x}} \, , \label{eq:imageforce}
\end{eqnarray}
where ${\bf \hat{x}}=(1,0,0)$ is an unit vector. 
Thus the vortex is attracted by the antivortex outside the film. 
It should be noted that an infinite number of images are required to satisfy the boundary condition.  
Under the assumption $d\gg \xi$, however, 
the vortex at $(x,y)=(\xi,0)$ receives a force from the nearest image at $(x,y)=(-\xi,0)$ dominantly, 
and contributions from other images are negligible.

In order to evaluate the force due to the external field, the distribution of screened field in the film is required.   
Since the magnetic field in the film can be derived from the London equation $d^2B/dx^2=B/\lambda^2$ with boundary conditions $B(0)=B_0$ and $B(d)=B_0$,  
we obtain $B(x)=B_0\cosh(\frac{x}{\lambda}-\frac{d}{2\lambda})/\cosh\frac{d}{2\lambda}$. 
Then the Meissner current at the vortex position is given by ${\bf j}_{\rm M}=(0,j_{{\rm M}\, y},0)$ and 
$j_{{\rm M}\,y} = -(1/\mu_0)(dB/dx)|_{x=\xi} \simeq (B_0/\mu_0\lambda)\tanh\frac{d}{2\lambda}$,  
where the assumption $d\gg \xi$ is used. 
Thus the force from the external field is given by 
\begin{eqnarray}
{\bf f}_{\rm M} 
= {\bf j}_{\rm M} \times \phi_0 {\bf \hat{z}} 
=  \frac{B_0\phi_0}{\mu_0\lambda} \tanh\frac{d}{2\lambda} {\bf \hat{x}}, \label{eq:externalforce}
\end{eqnarray}
by which the vortex is attracted to the inside of the film.

The force acting on the vortex is given by summation of the above two forces: ${\bf f}_{\rm tot} ={\bf f}_{\rm I}+{\bf f}_{\rm M}$. 
When the external field $B_0$ is so small that $|{\bf f}_{\rm I}| > |{\bf f}_{\rm M}|$, 
the force ${\bf f}_{\rm tot}$ directs the outside of the film. 
This force acts as a barrier that prevents the vortex penetration, which is called the Bean-Livingston barrier. 
When the external field $B_0$ is so large that $|{\bf f}_{\rm I}| < |{\bf f}_{\rm M}|$, the barrier disappears and the the vortex is drawn into the film. 
The external field that lets these two forces balance is called the vortex penetration field $B_v$, 
which can be evaluated by solving the equation %${\bf f}_{\rm tot}=0$. 
\begin{eqnarray}
{\bf f}_{\rm tot} =0 .  \label{eq:ftot}
\end{eqnarray}
When the film has an infinite thickness ($d\to \infty$), 
$B_v$ is reduced to the well-known expression for the semi-infinite superconductor, 
\begin{eqnarray}
B_v = \frac{\phi_0}{4\pi \lambda \xi} \simeq 0.7 B_c \,,  \label{Aeq:bvD}
\end{eqnarray}
where $B_c$ is the thermodynamic critical magnetic field. %which is identical to the well-known expression for the semi-infinite superconductor. 
On the other hand, when the film thickness is thinner than the London penetration depth ($d\ll\lambda$), 
$B_v$ is reduced to that for the thin film superconductor~\cite{stejic}, 
\begin{eqnarray}
B_v = \frac{\phi_0}{4\pi \lambda \xi} \frac{1}{\frac{d}{2\lambda}} = \frac{\phi_0}{2\pi d \xi}  \,. \label{eq:bvthin}
\end{eqnarray}
It should be noted that Eq.~(\ref{eq:bvthin}) can not be applied to the multilayer coating model, 
because the magnetic fields on the both sides of one superconductor layer have different amplitudes.  
In the next section, the vortex penetration field of the multilayer coating model is derived by reevaluating the force acting on the vortex.

%%%%%%%%%%%%%%%%%%%%%%%%%%%%%%%%%%%%%%%%%%%%%%%%%%%%%%%%%%%%%%%%%
%%%%%%%%%%%%%%%%%%%%%%%%%%%%%%%%%%%%%%%%%%%%%%%%%%%%%%%%%%%%%%%%%
\section{Vortex penetration field of the multilayer coating model}
%%%%%%%%%%%%%%%%%%%%%%%%%%%%%%%%%%%%%%%%%%%%%%%%%%%%%%%%%%%%%%%%%
%%%%%%%%%%%%%%%%%%%%%%%%%%%%%%%%%%%%%%%%%%%%%%%%%%%%%%%%%%%%%%%%%

%
\begin{figure}[tb]
   \begin{center}
   \includegraphics[width=0.8\linewidth]{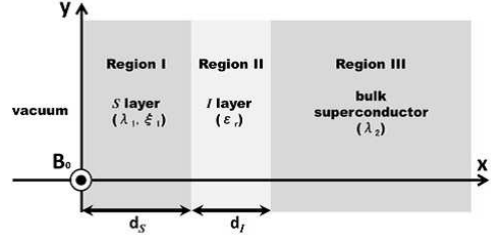}
   \end{center}\vspace{-0.6cm}
   \caption{
A multilayer coating model with a single $\mathcal{S}$ layer and a single $\mathcal{I}$ layer formed on a bulk superconductor.  
   }\label{figure2}
\end{figure}

Let us consider a model with a single $\mathcal{S}$ layer and a single $\mathcal{I}$ layer formed on a bulk superconductor as shown in Fig.~\ref{figure2}. 
The region $x<0$ is vacuum, 
the region I ($0 \le x \le d_{\mathcal{S}}$) is $\mathcal{S}$ layer with the London penetration depth $\lambda_1$, 
the region II ($d_{\mathcal{S}} < x < d_{\mathcal{S}} + d_{\mathcal{I}}$) is $\mathcal{I}$ layer with permittivity $\epsilon_r \epsilon_0$, 
in which $\epsilon_r$ is a relative permittivity,   
and the region III ($x \ge d_{\mathcal{S}} +d_{\mathcal{I}}$) is a bulk superconductor with the London penetration depth $\lambda_2$, 
where all layers are parallel to the $y$-$z$ plane and then perpendicular to the $x$-axis.  
The applied electric and magnetic field are assumed to be parallel to the layers.

The force due to the image vortex ${\bf f}_{\rm I}$ is given by Eq.~(\ref{eq:imageforce})  
when the material of the $\mathcal{S}$ layer and its thickness satisfy $\lambda_1 \gg \xi_1$ and $d_{\mathcal{S}} \gg \xi_1$.  
In order to evaluate the force due to the external magnetic field ${\bf f}_{\rm M}$, the magnetic field distribution in the $\mathcal{S}$ layer is required,  
which can be derived by solving the London equations in the region I and III, and the Maxwell equations in the region II, 
with boundary conditions given as continuity conditions of the electric and magnetic field at $x=d_{\mathcal{S}}$ and $x=d_{\mathcal{S}}+d_{\mathcal{I}}$~\cite{ipac13kis}. 
Assuming an insulator thickness $d_{\mathcal{I}} \ll (\sqrt{\epsilon_r}k)^{-1}$,  
the magnetic field attenuation formula for the region I can be reduced to the simple form~\cite{kis}:  
\begin{eqnarray}
B = 
B_0
\frac{ \lambda_1 \cosh \frac{d_{\mathcal{S}}-x}{\lambda_1} + (\lambda_2 +d_{\mathcal{I}}) \sinh \frac{d_{\mathcal{S}}-x}{\lambda_1}}
{\lambda_1 \cosh \frac{d_{\mathcal{S}}}{\lambda_1} + (\lambda_2 + d_{\mathcal{I}}) \sinh \frac{d_{\mathcal{S}}}{\lambda_1}} \, .    
\label{eq:BI} 
\end{eqnarray}
It should be noted that Eq.~(\ref{eq:BI}) is reduced to the well known expression $B=B_0 e^{-x/\lambda_1}$, 
when the ${\mathcal{S}}$ layer thickness is infinite ($d_{\mathcal{S}}\to\infty$). 
Then the Meissner current in the ${\mathcal{S}}$ layer, $j_{{\rm M}y}=-(1/\mu_0)dB/dx$, is given by 
\begin{eqnarray}
j_{{\rm M}y}  
= \frac{B_0}{\mu_0 \lambda_1}
\frac{ \lambda_1 \sinh \frac{d_{\mathcal{S}}-x}{\lambda_1} + (\lambda_2 +d_{\mathcal{I}}) \cosh \frac{d_{\mathcal{S}}-x}{\lambda_1}}
{\lambda_1 \cosh \frac{d_{\mathcal{S}}}{\lambda_1} + (\lambda_2 + d_{\mathcal{I}}) \sinh \frac{d_{\mathcal{S}}}{\lambda_1}} \,, 
\label{eq:jMy}
\end{eqnarray}
and thus the Lorentz force, ${\bf f}_{\rm M} = {\bf j}_{\rm M} \times \phi_0 {\bf \hat{z}}$, is given by 
\begin{eqnarray}
{\bf f}_{\rm M} = \frac{B_0\phi_0}{\mu_0 \lambda_1}
\frac{ \lambda_1 \sinh \frac{d_{\mathcal{S}}-x}{\lambda_1} + (\lambda_2 +d_{\mathcal{I}}) \cosh \frac{d_{\mathcal{S}}-x}{\lambda_1}}
{\lambda_1 \cosh \frac{d_{\mathcal{S}}}{\lambda_1} + (\lambda_2 + d_{\mathcal{I}}) \sinh \frac{d_{\mathcal{S}}}{\lambda_1}} {\bf \hat{x}} 
\,. \label{eq:fM_multilayer}
\end{eqnarray}
Balancing the Lorentz forces given by Eq.~(\ref{eq:imageforce}) and Eq.~(\ref{eq:fM_multilayer}),  
we obtain the vortex penetration field of the multilayer coating model. 
Completed results and discussions are seen in Ref.~\cite{kis}.

It should be noted that 
the resultant formula can be applied to a model with an ${\mathcal{S}}$ layer formed on a bulk superconductor without an ${\mathcal{I}}$ layer ($d_{\mathcal{I}}\to 0$)~\cite{kis}.

%%%%%%%%%%%%%%%%%%%%%%%%%%%%%%%%%%%%%%%%%%%%%%%%%%%%%%%%%%%%%%%%%
%%%%%%%%%%%%%%%%%%%%%%%%%%%%%%%%%%%%%%%%%%%%%%%%%%%%%%%%%%%%%%%%%
\section{Summary}
%%%%%%%%%%%%%%%%%%%%%%%%%%%%%%%%%%%%%%%%%%%%%%%%%%%%%%%%%%%%%%%%%
%%%%%%%%%%%%%%%%%%%%%%%%%%%%%%%%%%%%%%%%%%%%%%%%%%%%%%%%%%%%%%%%%

The vortex penetration field of the multilayer coating model was derived from the RF electromagnetic field attenuation formulae~\cite{kis, ipac13kis}. 
The same formula can be applied to 
a model with a single ${\mathcal{S}}$ layer formed on a bulk superconductor without an ${\mathcal{I}}$ layer.

\end{document}